\documentclass[twocolumn,superscriptaddress,amsmath,amssymb,aps,prb,longbibliography]{revtex4-2}

\usepackage[dvipsnames]{xcolor}
\usepackage{graphicx}% Include figure files
\usepackage{dcolumn}% Align table columns on decimal point
\usepackage{bm}% bold math
\usepackage[colorlinks, breaklinks, linkcolor=OrangeRed, citecolor=RoyalBlue, urlcolor=NavyBlue]{hyperref}
\usepackage{siunitx}
\usepackage[normalem]{ulem}
\newcommand{\new}[1]{\textcolor{black}{#1}}
\usepackage{nicefrac}
\usepackage{xspace}
\usepackage{float}
\usepackage{mathbbol}
\usepackage[utf8]{inputenc}
\DeclareSymbolFontAlphabet{\mathbb}{AMSb}     % AMS bold is prettier ...
\DeclareSymbolFontAlphabet{\mathbbl}{bbold}   % but needs more symbols.

\begin{document}
\title{Comment on ``Absence versus Presence of Dissipative Quantum Phase Transition in Josephson Junctions''}

\author{Théo Sépulcre}
\affiliation{Wallenberg Centre for Quantum Technology, Chalmers University of Technology, SE-412 96 Gothenburg, Sweden}
\author{Serge Florens}
\affiliation{Univ. Grenoble Alpes, CNRS, Grenoble INP, Institut Néel, 38000 Grenoble, France}
\author{Izak Snyman}
\affiliation{Mandelstam Institute for Theoretical Physics, School of Physics, University of the Witwatersrand,
Johannesburg, South Africa}
\email{izak.snyman@wits.ac.za}

\maketitle

In Ref. \cite{Masuki2022}, a Josephson
junction shunted by an ohmic transmission line is studied. 
%with Josephson energy $E_\text{J}$ and charging energy $E_\text{C}$, 
%with conductance $\alpha (2e)^2/h$. 
%Their model includes a realistic high 
%frequency cutoff of order $\alpha E_c$, that is typically smaller than the plasma frequency $W$.
The authors present a phase diagram with features not anticipated in the established 
literature~\cite{Schoen_Review}.
%For $E_\text{J}/E_\text{C}$ above a certain value, they find that the junction remains superconducting for all 
%$\alpha$, while below this value, they find that the insulating phase leads to re-entrant superconductivity at 
%small $\alpha$. 
We show that their Numerical Renormalization Group (NRG) calculation suffers from several flaws, and 
cannot be trusted to substantiate their claims.
%that there
%is no evidence for the re-entrant superconductivity in the phase diagram presented in Fig.\,1a of
%\cite{Masuki2022}. 

%An essential requirement is scale separation, which entails 
%that $\Delta H_{N+1}$ should constitute a perturbation to $H_N$ whose
%strength decreases exponentially fast at increasing $N$. For instance, in Wilson's treatment of
%the Kondo problem, $\Delta H_{N+1}=t_{N+1}\sum_{\sigma=\pm}(c_{N\pm}^\dagger c_{N+1\pm}+\text{h.c})$
%with $t_N\sim \Lambda^{-N/2}$, $\Lambda>1$, and $\{c_N\}$ a set of fermionic annihilation operators.
%This requirement is referred to as scale separation, and allows one to obtain accurate ground state
%properties of $\lim_{N\to\infty} H_N$ while only retaining a fixed number of low energy states of
%$H_N$ at each step.
\begin{figure}[ht!]
\begin{center} \includegraphics[width = 0.8\columnwidth]{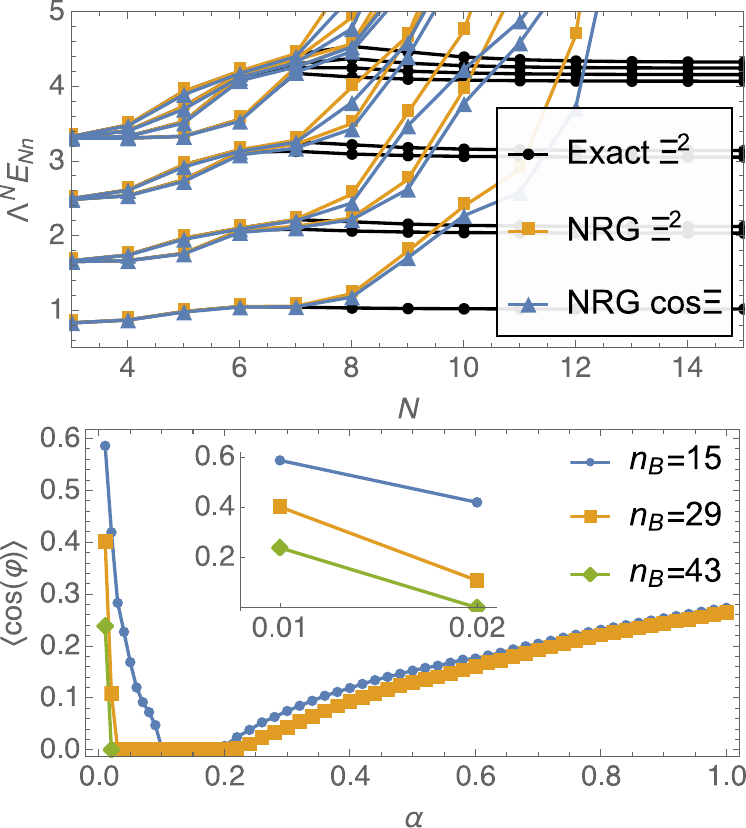}
\caption{{\bf Top}: Low energy spectrum v. NRG step $N$, scaled with $\Lambda^N$.
Results of the NRG scheme in \cite{Masuki2022} for the cosine and quadratic potential are
compared to exact results for the quadratic potential. We took $n_S=50$ kept states, 
$n_B=300$ bosonic states for $N=0$ and $n_B=15$ for $N>0$, $\Lambda=2.0$, $\alpha=10$, $E_\text{C}=0.01W$, 
$E_{\text J}/E_\text{C}=10$. 
{\bf Bottom}: $\left<\cos(\varphi)\right>$ v. $\alpha$, for
$E_\text{J}/E_\text{C}=0.15$, like the triangles in the top panel of Fig.~4 of \cite{Masuki2022}.
The blue dots reproduce the result of \cite{Masuki2022} with the same truncation parameter 
$n_B=15$ for $N>0$.
The yellow squares and green diamonds were obtained by increasing $n_B$ to $29$ and $43$ respectively. 
The inset zooms in on the two smallest values of $\alpha$, which are still unconverged at $n=43$,
showing a downward trend.}
\label{fig1}
\end{center}
\end{figure}

NRG captures low energy physics by building recursive Hamiltonians,
%through the recursion 
$H_{N+1}=H_N+\Delta H_{N+1}$, that are iteratively diagonalized.
Scale separation is required for NRG to work, i.e. $\Delta H_{N+1}$
should decrease exponentially with $N$~\cite{Bulla}. For the NRG scheme in
Ref.~\cite{Masuki2022}, $\Delta H_{N+1}$ is
of the same order as $H_0$ 
[See Eqs. (S51) and (S52) in the supplementary material to \cite{Masuki2022}.]. This is a known problem 
that can only be cured by introducing an infrared cut-off \cite{Freyn2011}.
As a result, the NRG fails to flow to the correct infrared fixed point.   
%should couple the system to new degrees of freedom at successively lower energy scales.
%In Ref.~\cite{Masuki2022}, 
%$\Delta H_{N+1}$ contains 
%an order $\Lambda^0$ 
%a term
%$-E_\text{J}\left(\cos\Xi_{N+1}-\cos\Xi_N\right)$ where mode $a_{N+1}$ contributes to the phase variable
%$\Xi_{N+1}=\Xi_N+i\xi_{N+1}(a_{N+1}^\dagger-a_{N+1})$ with the amplitude
%$\xi_{N+1}\simeq2\sqrt{(\Lambda-1)/(\Lambda+1)}/\sqrt{\alpha}$ at large $N$. This violates scale separation,
%which demands that the strength of $\Delta H_{N}$ decreases exponentially with $N$, for NRG to work \cite{Bulla}. 
%That the NRG scheme fails, is evident from the $\sim \Lambda^N$ growth of the low-energy spectrum of $\Lambda^N H_N$ in the right panel
%of Fig.\,S2 in the supplementary material of \cite{Masuki2022}. 
%All excitation energies  of
%$\Lambda^N H_N$, computed by NRG, start growing like $\Lambda^N$ after $N=30$. 
%The spectrum of $H_N$
%obtained in NRG 
%thus 
%saturates after $N=30$. In contrast, the true spectrum of $H_N$ has excitation
%energies down to $\Lambda^{-N}$ for arbitrarily large $N$.
%, due to low frequency modes.
%To investigate this further, 
To demonstrate this, we considered
large conductance $\alpha$ and large $E_\text{J}/E_\text{C}$, where the system studied in \cite{Masuki2022} is nearly harmonic, 
allowing us to expand $-E_\text{J}\cos(\Xi)\simeq E_\text{J}(\Xi^2/2-1)$. 
%in order to gain analytic insights.
We compared low energy spectra obtained with the NRG scheme of \cite{Masuki2022} for the cosine and quadratic 
potentials, to the exact spectrum obtained for the latter. 
%If the NRG scheme is well-behaved, the three 
%calculations should yield nearly the same result. However, 
As the top panel of Fig.\,\ref{fig1} shows, the NRG 
results diverge from the exact spectrum after the seventh RG step.
%\new{Unsurprisingly, the same problems are found when examining the tails of the mobility
%$\mu_{10}$, invalidating recent claims~\cite{Reply}.}
%, in the same manner 
%as the exponential flow seen in the right panel of Fig.\,S2 in the supplementary material accompanying
%\cite{Masuki2022}. 
Thus the NRG scheme proposed in \cite{Masuki2022} is unreliable and cannot be trusted to predict the phase 
diagram. (See Appendix \ref{appa} for discussion of the RG flow of mobility $\mu_{10}$.)

%We must further point out that, 
The phase diagram in \cite{Masuki2022} is flawed in another way. Even if one  
trusted the employed NRG scheme, the re-entrant superconductivity seen at small $\alpha$ and
small $E_\text{J}/E_\text{C}$ is a numerical artefact. The blue dots in the bottom panel of
Fig.\,\ref{fig1} reproduce the result for $\left<\cos(\varphi)\right>$ v. $\alpha$ at 
$E_\text{J}/E_\text{C}=0.15$ in the upper panel of Fig.\,4 of \cite{Masuki2022}, obtained 
with the truncation parameter $n_B=15$ in each mode for $N>0$.
%In \cite{Masuki2022}, the region between $\alpha=0.1$ and
%$\alpha=0.2$, where $\left<\cos(\varphi)\right>$  vanishes, is identified as the insulating phase.
%In the exact Hamiltonian, there is no sharp cut-off on the number of bosons allowed in any mode. 
For this result to be correct, it must not change when $n_B$ is increased. Instead we see that the
region where $\left<\cos(\varphi)\right>$ vanishes, grows to include the interval $\alpha\in[0,0.2]$
when $n_B$ is increased. Thus, the apparent re-entrant superconductivity in
the phase diagram in \cite{Masuki2022} stems from unconverged data. 
In \cite{Masuki2022} it is argued that superconductivity makes common sense when 
the junction is shunted by a sufficiently large impedance. 
We stress
that taking the thermodynamic limit $N\to \infty$ before $\alpha\to0$,
couples the junction to divergent $\varphi$-fluctuations that render the junction's zero-frequnecy response non-trivial.
The object Letter also contains a brief functional Renormalization Group (fRG) argument in support 
of superconductivity at $\alpha<1$ and large $E_\text{J}/E_\text{C}$. 
The approximations involved are not controlled by any obvious small parameter. It is still not known whether fRG can reproduce infrared Luttinger exponents for
$1<\alpha<2$~\cite{Freyn2011}, where phase-slips affect results non-trivially. Until this is settled, fRG's validity in the more challenging $\alpha<1$ regime remains unclear.

\appendix

\section{Additional Information}
\label{appa}
Here we present further information that length restrictions prevented us from presenting in the published 
comment. It concerns the contribution
$\mu_{10}$ to the phase mobility, that is employed as an order parameter in the Object Letter. 
\begin{figure}[H]
\begin{center}
\includegraphics[width = 0.99\columnwidth]{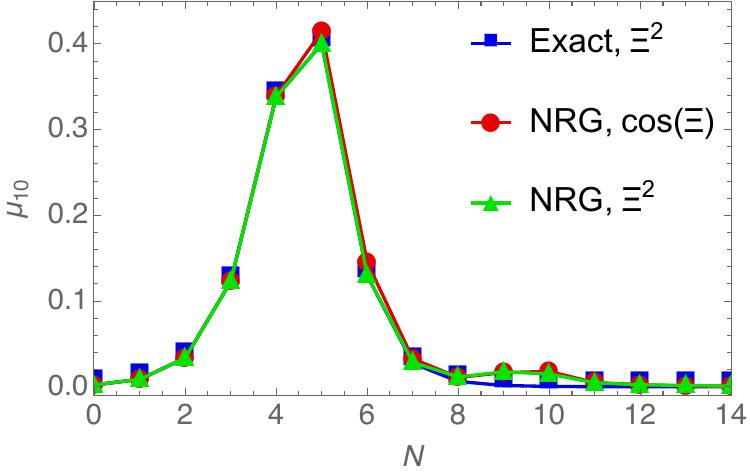}
\includegraphics[width = 0.99\columnwidth]{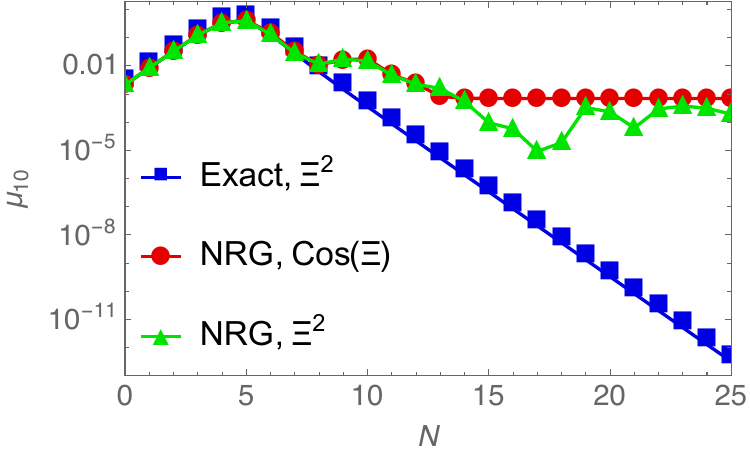}
\caption{ {\bf Top panel:} Mobility $\mu_{10}$ \new{as a function of site index $N$ in the NRG
discretization}, for $E_\text{J}=10 E_\text{C}$, $\alpha=10$,
$N_\text{kept}=50$, $n_B=14$ for $n>0$. Other parameters as in the Comment. 
{\bf Bottom panel:} Same quantity, replotted on a logarithmic vertical axis, showing the breakdown
in the NRG computation, both for the cosine (red dots) and quadratic (green triangles) potentials.
In contrast to the vanishing mobility that is correctly obtained from the exact result (blue
squares) in the superconducting phase, the NRG leads to a finite mobility at the end of the flow
(namely $N\gg1)$.}
\label{figa1}
\end{center}
\end{figure}
We have calculated $\mu_{10}$, which sheds further light on the convergence issues pointed
out in the Comment, \new{and investigated is dependence with $N$, the number of sites in the NRG
discretization. This observable shows} a crossover between ultraviolet behaviour
(small $N$) and infrared behavior (large $N$). At small $N$, one has:
\begin{equation}
\mu_{10}\simeq \alpha \xi_N^2.
\end{equation}
\new{(See the Supplemental Material to the Object Letter for details on the notation).}
In the harmonic limit, where the cosine potential can be replaced by one that is quadratic, the
asymptotic behavior at large $N$ is
\begin{equation}
\mu_{10}\simeq \left(\frac{\Lambda}{2}\right)^4\frac{\gamma_N^2}{8 E_\text{J}^2\xi_N^4}.
\end{equation}
The top panel of Fig.~\ref{figa1} shows behavior very similar to the non-monotone flow presented in
Fig.~3 of the Object Letter. It shows the flow of the mobility, which at first sight seems to
indicate that the NRG results are reliable and lead to a vanishing mobility in the ground state.
However, the correct value of the mobility $\mu_{10}$ should be read after complete iteration of the
NRG scheme (namely for large $N$ values in the plot), corresponding to the final stage of the
renormalization flow. In the bottom panel of Fig.~\ref{figa1}, we therefore show exactly the same
mobility as in the top panel, but we have extended the horizontal axis to larger $N$ and displayed
the data in a \new{better} way using a logarithmic scale. What can be seen here is again a complete
breakdown of this NRG after few iterations: the mobility does not vanish (either for the cosine or
quadratic potentials), contrary to the claim of Masuki {\it et al.}. Rather, the mobility saturates
to a finite value in the NRG simulation, which is physically incorrect. In contrast, the exact
solution does display a vanishing mobility at large $N$, as expected in the superconducting phase of
the model. The same issues are found for all values of the parameters of the considered model, and
thus the results cannot be trusted to establish a phase diagram. Again, we stress that these
problems are fully expected since the NRG of Masuki {\it et al.} does not converge. A devil's
advocate could perhaps argue that a finite but small mobility could be used as an approximate way to
describe the superconducting phase, although there would be no qualitative difference with the
insulating regime, so that a careful scaling analysis would be required to establish a proper phase
diagram. However, this argument cannot be made, because all the NRG calculations of Masuki {\it et
al.} are plagued by convergence problems. A clear example for this serious issue is given for the
parameters $E_\text{J}=10 E_\text{C}$, $\alpha=2$, which indisputably fall inside the
superconducting phase. If we increase the truncation parameters $n_B$ and $N_\text{kept}$,
from respectively $15$ and $50$ (their values in the original Letter by Masuki {\it et al.}) to 
respectively $29$ and $100$, which should make the result more accurate, the mobility $\mu_{10}$ 
switches from superconducting-like to strongly insulating for the cosine potential, see Fig.~\ref{fig2} 
below.

A further point, of importance to anyone wishing to reproduce our results, or those of the object
Letter, is the following: To achieve agreement with the results in the Object Letter, we had to
reverse engineer a mistake in the numerics, whose presence is revealed by the spectra of
Fig.\protect \tmspace +\thinmuskip {.1667em}S2 in the Supplemental Material to the Object Letter.
At low $N$, the spectrum should be $n\gamma _0$, $n=1,2,\protect \ldots $. According to Eq.~(S50), 
one should have $\gamma _0=4.33$ in Fig.\protect \tmspace
+\thinmuskip {.1667em}S2. Instead, one sees $\gamma _0=0.5$. There is in fact a typo in Eq.\protect
\tmspace +\thinmuskip {.1667em}(S50): $(1+\Lambda +\Lambda ^{-2})/3$ should be replaced with
$(1+\Lambda ^{-1}+\Lambda ^{-2})/3$. Correcting the typo however still does not give $\gamma _0=0.5$ as
in Fig.\protect \tmspace +\thinmuskip {.1667em}S2. We therefore tried various plausible mistakes,
and we found one that reproduced exactly the NRG results in the Object Letter. Apparently, the
numerics in the Object Letter were performed using $(1+\Lambda ^{-3})/3$ instead of $(1+\Lambda
+\Lambda ^{-2})/3$ in Eq.\protect \tmspace +\thinmuskip {.1667em}(S50) that defines $\gamma _n$.
When the correct expression for $\gamma _n$ is used, the position of the phase boundary in the
phase diagram changes, \new{making the apparent agreement to fRG predictions rather fortuitous}. 

\begin{figure}[H]
\begin{center}
\includegraphics[width = 0.99\columnwidth]{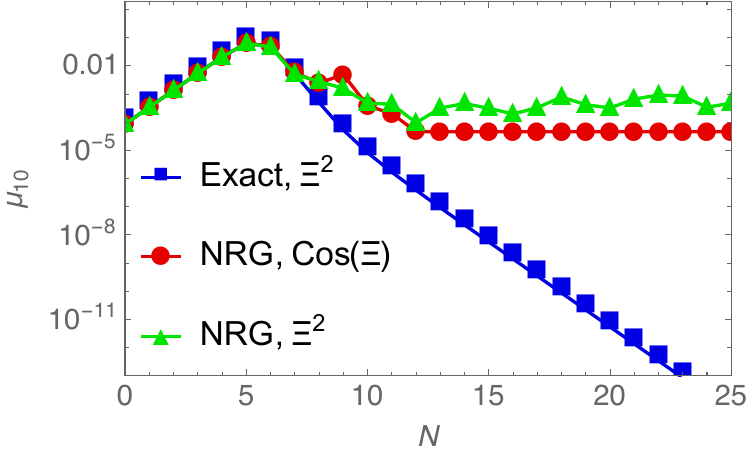}
\includegraphics[width = 0.99\columnwidth]{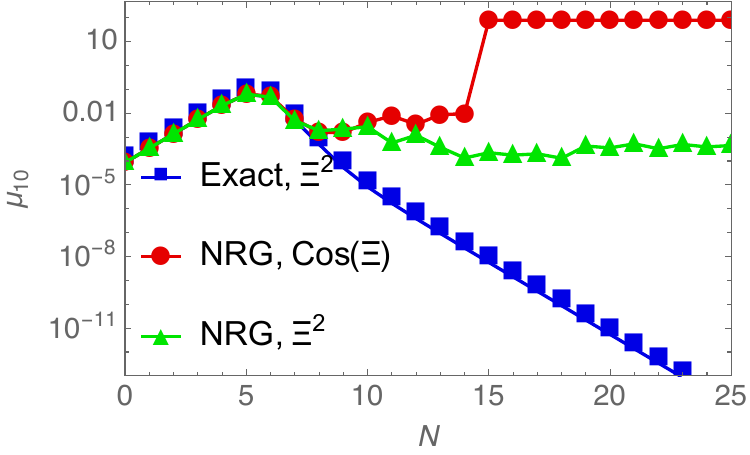}
\caption{Convergence issues in the NRG scheme of Masuki {\it et al.}, comparing 
two choices of truncation parameters.
{\bf Top panel:} $N_\text{kept}=50$, $n_B=14$ for $n>0$, leading to a finite but small
mobility.
{\bf Bottom panel:} 
$E_\text{J}=N_\text{kept}=100$, $n_B=29$ for $n>0$, leading to a finite but large mobility.
The mobility should in any case vanish in the limit $N\gg1$, since parameters
$E_\text{J}= 10 E_\text{C}$, $\alpha=2$ correspond to the superconducting phase.
}
\label{fig2}
\end{center}
\end{figure}

%the re-entrant superconductivity claimed in \cite{Masuki2022} at small $\alpha$ 
%goes away when a sufficiently large number of bosonic excitations per mode
%is allowed in the NRG scheme.
%While related Josephson Hamiltonians displaying spin-boson physics are immune to these
%limitations~\cite{Kaur}, calculating ground state properties of the boundary sine Gordon model using 
%NRG remains challenging. 

%apsrev4-2.bst 2019-01-14 (MD) hand-edited version of apsrev4-1.bst
%Control: key (0)
%Control: author (8) initials jnrlst
%Control: editor formatted (1) identically to author
%Control: production of article title (0) allowed
%Control: page (0) single
%Control: year (1) truncated
%Control: production of eprint (0) enabled
%


\begin{thebibliography}{a}%
\makeatletter
\providecommand \@ifxundefined [1]{%
 \@ifx{#1\undefined}
}%
\providecommand \@ifnum [1]{%
 \ifnum #1\expandafter \@firstoftwo
 \else \expandafter \@secondoftwo
 \fi
}%
\providecommand \@ifx [1]{%
 \ifx #1\expandafter \@firstoftwo
 \else \expandafter \@secondoftwo
 \fi
}%
\providecommand \natexlab [1]{#1}%
\providecommand \enquote  [1]{``#1''}%
\providecommand \bibnamefont  [1]{#1}%
\providecommand \bibfnamefont [1]{#1}%
\providecommand \citenamefont [1]{#1}%
\providecommand \href@noop [0]{\@secondoftwo}%
\providecommand \href [0]{\begingroup \@sanitize@url \@href}%
\providecommand \@href[1]{\@@startlink{#1}\@@href}%
\providecommand \@@href[1]{\endgroup#1\@@endlink}%
\providecommand \@sanitize@url [0]{\catcode `\\12\catcode `\$12\catcode
  `\&12\catcode `\#12\catcode `\^12\catcode `\_12\catcode `\%12\relax}%
\providecommand \@@startlink[1]{}%
\providecommand \@@endlink[0]{}%
\providecommand \url  [0]{\begingroup\@sanitize@url \@url }%
\providecommand \@url [1]{\endgroup\@href {#1}{\urlprefix }}%
\providecommand \urlprefix  [0]{URL }%
\providecommand \Eprint [0]{\href }%
\providecommand \doibase [0]{https://doi.org/}%
\providecommand \selectlanguage [0]{\@gobble}%
\providecommand \bibinfo  [0]{\@secondoftwo}%
\providecommand \bibfield  [0]{\@secondoftwo}%
\providecommand \translation [1]{[#1]}%
\providecommand \BibitemOpen [0]{}%
\providecommand \bibitemStop [0]{}%
\providecommand \bibitemNoStop [0]{.\EOS\space}%
\providecommand \EOS [0]{\spacefactor3000\relax}%
\providecommand \BibitemShut  [1]{\csname bibitem#1\endcsname}%
\let\auto@bib@innerbib\@empty
%</preamble>
\bibitem [{\citenamefont {Masuki}\ \emph {et~al.}(2022)\citenamefont {Masuki},
  \citenamefont {Sudo}, \citenamefont {Oshikawa},\ and\ \citenamefont
  {Ashida}}]{Masuki2022}%
  \BibitemOpen
  \bibfield  {author} {\bibinfo {author} {\bibfnamefont {K.}~\bibnamefont
  {Masuki}}, \bibinfo {author} {\bibfnamefont {H.}~\bibnamefont {Sudo}},
  \bibinfo {author} {\bibfnamefont {M.}~\bibnamefont {Oshikawa}},\ and\
  \bibinfo {author} {\bibfnamefont {Y.}~\bibnamefont {Ashida}},\ }\bibfield
  {title} {\bibinfo {title} {Absence versus presence of dissipative quantum
  phase transition in josephson junctions},\ }\href
  {https://doi.org/10.1103/PhysRevLett.129.087001} {\bibfield  {journal}
  {\bibinfo  {journal} {Phys. Rev. Lett.}\ }\textbf {\bibinfo {volume} {129}},\
  \bibinfo {pages} {087001} (\bibinfo {year} {2022})}\BibitemShut {NoStop}%
\bibitem [{\citenamefont {Sch{\"o}n}\ and\ \citenamefont
  {Zaikin}(1990)}]{Schoen_Review}%
  \BibitemOpen
  \bibfield  {author} {\bibinfo {author} {\bibfnamefont {G.}~\bibnamefont
  {Sch{\"o}n}}\ and\ \bibinfo {author} {\bibfnamefont {A.~D.}\ \bibnamefont
  {Zaikin}},\ }\bibfield  {title} {\bibinfo {title} {{Quantum coherent effects,
  phase transitions, and the dissipative dynamics of ultra small tunnel
  junctions}},\ }\href
  {https://doi.org/https://doi.org/10.1016/0370-1573(90)90156-V} {\bibfield
  {journal} {\bibinfo  {journal} {Physics Reports}\ }\textbf {\bibinfo {volume}
  {198}},\ \bibinfo {pages} {237} (\bibinfo {year} {1990})}\BibitemShut
  {NoStop}%
\bibitem [{\citenamefont {Bulla}\ \emph {et~al.}(2008)\citenamefont {Bulla},
  \citenamefont {Costi},\ and\ \citenamefont {Pruschke}}]{Bulla}%
  \BibitemOpen
  \bibfield  {author} {\bibinfo {author} {\bibfnamefont {R.}~\bibnamefont
  {Bulla}}, \bibinfo {author} {\bibfnamefont {T.~A.}\ \bibnamefont {Costi}},\
  and\ \bibinfo {author} {\bibfnamefont {T.}~\bibnamefont {Pruschke}},\
  }\bibfield  {title} {\bibinfo {title} {Numerical renormalization group method
  for quantum impurity systems},\ }\href
  {https://doi.org/10.1103/RevModPhys.80.395} {\bibfield  {journal} {\bibinfo
  {journal} {Rev. Mod. Phys.}\ }\textbf {\bibinfo {volume} {80}},\ \bibinfo
  {pages} {395} (\bibinfo {year} {2008})}\BibitemShut {NoStop}%
\bibitem [{\citenamefont {Freyn}\ and\ \citenamefont
  {Florens}(2011)}]{Freyn2011}%
  \BibitemOpen
  \bibfield  {author} {\bibinfo {author} {\bibfnamefont {A.}~\bibnamefont
  {Freyn}}\ and\ \bibinfo {author} {\bibfnamefont {S.}~\bibnamefont
  {Florens}},\ }\bibfield  {title} {\bibinfo {title} {Numerical renormalization
  group at marginal spectral density: Application to tunneling in luttinger
  liquids},\ }\href {https://doi.org/10.1103/PhysRevLett.107.017201} {\bibfield
   {journal} {\bibinfo  {journal} {Phys. Rev. Lett.}\ }\textbf {\bibinfo
  {volume} {107}},\ \bibinfo {pages} {017201} (\bibinfo {year}
  {2011})}\BibitemShut {NoStop}%
\end{thebibliography}
\end{document}